\documentclass[12pt]{article}
\font\amsmathe=msbm10 scaled \magstep1

\def\N{\hbox{\amsmathe N}}
\def\Q{\hbox{\amsmathe Q}}
\def\R{\hbox{\amsmathe R}}

\begin{document}                                                                
\begin{titlepage}
\title{New Universality Classes\\
in One--Dimensional $O(N)$--Invariant Spin--Models\\
with an $n$--Parametric Action}

\author{{\bf Erhard Seiler} \\ \\ 
Max--Planck--Institut f\"ur Physik\\
--- Werner--Heisenberg--Institut ---\\
F\"ohringer Ring 6, D--80805 M\"unchen, Germany\\
\\
and
\\[1cm]
{\bf Karim Yildirim}\\  \\
Max--Planck--Institut f\"ur Physik\\
--- Werner--Heisenberg--Institut ---\\
F\"ohringer Ring 6, D--80805 M\"unchen, Germany}

\maketitle
\vspace{-16cm}
\begin{flushright} {\bf MPI-PTh/96-8}
\end{flushright}
\thispagestyle{empty}
\vspace{18cm}

\begin{abstract}
An action with $n$ parameters, which generalizes the
$O(N)$--$R P^{N-1}$--model, is considered in one dimension for general
$N$. We use asymptotic expansion techniques to determine where the model 
becomes critical and show that for the actions considered
there exists a family of hypersurfaces whose asymptotic behaviour 
determines a one--parameter family of new universality
classes. They interpolate between the
$O(N)$--vector--model--class and the $R P^{N-1}$--model--class. Furthermore
continuum limits are discussed, including the exceptional case $N=2$.
\end{abstract}

\end{titlepage}
\newpage

\section[Introduction]{Introduction}

The question of universality is central for lattice field theories.
It is generally tacitly assumed that it does not matter which lattice
discretization of a classical action is employed, if one is interested 
only in the continuum limit. Recently Caracciolo et al.\ \cite{Caracciolo} have
sown doubt about this universality for the two--dimensional ($2D$) $O(N)$
nonlinear $\mbox{\boldmath$\sigma$}$--models. Their claim that there are 
different universality classes once one introduces in addition to the standard
`isovector' coupling an `isotensor' term in the action has generated
some controversy. Both Niedermayer et al.\ \cite{NSW} and Hasenbusch
\cite{Hasenbusch} argue that this violation of the universality dogma is
only apparent and that it disappears as soon as one is looking at the right
observables.

Since a definite mathematical answer to this question for the $2D$ models
is out of reach, it seems worthwhile to study the question in the exactly
solvable $1D$ model. While we were working on this question, Cucchieri et
al.\ \cite{Sokal} produced a lengthy
paper on the subject; their conclusions agree to
a large extent with our findings. But we find that their paper does not
answer all the questions that come to mind. For instance they study 
mostly one--parameter families of rather general coupling functions as
well as a 2--parameter family, but in much less detail and generality.

In this paper we consider $n$--parameter families of actions that are
natural gene\-ralization of the actions studied in \cite{Caracciolo}. 
We still obtain only a one--parameter family of universality classes, just
as in the cases examined in \cite{Sokal}. On the other hand it turns out 
that these different universality classes reflect the true spectral 
properties of the transfer matrix, whereas the reinterpretations proposed
by \cite{NSW} and \cite{Hasenbusch}, which reduce everything to the 
`standard' universality class are unrelated to the transfer matrix.

This paper is organized as follows : in section \ref{Preliminaries} we 
introduce an $n$-parameter family of actions for $1D$
$O(N)$--invariant spin models taking values on the sphere $S^{N-1}$, with 
nearest--neighbour interactions. It generalizes the generic mixed 
isovector/isotensor--model. The main result is then presented in section 
\ref{MainResult} : using asymptotic expansion techniques we find where 
and in which way the
models become critical. Especially, there are hypersurfaces on which an
infinite number of new universality classes appear. In the next section
it is shown that the restrictions on the Hamiltonian made in section
\ref{MainResult} are not essential (in the case of {\it non--negative} 
parameters). In section \ref{contlim} we also discuss the continuum
limit and give a supplement to the paper \cite{Sokal}. Finally, our 
conclusions are stated in section \ref{conclusion}.

\section[Preliminaries]{Preliminaries\label{Preliminaries}}

We want to study the critical behaviour of spin models which are
generalizations of the well--known $O(N)$--$R P^{N-1}$--model
\cite{Caracciolo}. Therefore, we consider nearest--neighbour interactions 
given by
polynomials $\sum\limits_{k=1}^n\beta_kx^k$ in the $O(N)$--invariant
scalar product $\mbox{\boldmath$\sigma$}\cdot\mbox{\boldmath$\sigma'$}$, i.\ e.\
\begin{equation}
{\cal S}:=\sum\limits_x\sum\limits_{k=1}^n \beta_k \left(\mbox{
\boldmath$\sigma$}_x\cdot\mbox{\boldmath$\sigma$}_{x+1}\right)^k.
\label{action}
\end{equation}
The spin $\mbox{\boldmath$\sigma$}$ takes values on the sphere $S^{N-1}\subset
\R^N$ (with the $O(N)$--invariant, normalized measure
$d\Omega(\mbox{\boldmath$\sigma$})$) and all parameters 
$\mbox{\boldmath$\beta$}=(\beta_1,\ldots,\beta_n)\in\R^n_{\ge 0}$
are nonnegative.
The Hilbert space $L^2(S^{N-1})$ can be decomposed into
the eigenspaces ${\cal H}_l$ of the {\it Laplace--Beltrami--operator}
$\Delta_{LB}$, corresponding to the eigenvalues $-l(l+N-2)$; since
$\Delta_{LB}$ is a Casimir element of the Lie algebra of
$O(N)$, these
eigenspaces are invariant subspaces under the right action of $O(N)$.
The projections onto the corresponding eigenspaces are given by the
integral kernels
\begin{equation}
{\cal P}_l(\mbox{\boldmath$\sigma$},\mbox{\boldmath$\sigma'$})
={2l+N-2\over N-2} C_l^{N-2\over 2}
(\mbox{\boldmath$\sigma$}\cdot\mbox{\boldmath$\sigma'$})
\label{projectors}
\end{equation}
where the integrals are to be taken with the measure 
$d\Omega(\mbox{\boldmath$\sigma$})$, and $C^{\frac{N}{2}-1}_l$ are the
{\it Gegenbauer--polynomials} \cite{Magnus}.

The transfer matrix corresponding to (\ref{action}) is given by the 
integral kernel
\begin{equation}
{\cal T}(\mbox{\boldmath$\sigma$}\cdot\mbox{\boldmath$\sigma'$}):=
\mbox{exp}\left(\sum_{k=1}^n \beta_k(\mbox{\boldmath$\sigma$}\cdot\mbox{
\boldmath$\sigma'$})^k\right).
\end{equation}
Because $\cal T$ commutes with the $O(N)$ rotations, it will 
also leave these eigenspaces invariant, and in fact it will act as 
multiplication by the eigenvalue $\lambda_l$ on these subspaces.
Hence we can evaluate the eigenvalues $\lambda_l$ for
$N\in\N\backslash\{1\}$ as
\begin{eqnarray}
\lambda_l(\mbox{\boldmath$\beta$})&:=&
    \frac{\mbox{tr}{\cal P}_l{\cal T}}{\mbox{tr}{\cal P}_l}\nonumber\\
&=&\int\limits^1_{-1} \mbox{exp}\left(\sum\limits^n_{k=1} 
\beta_k x^k\right)\left(1-x^2\right)^{\frac{N-3}{2}}
\frac{C_l^{\frac{N}{2}-1}(x)}{C_l^{\frac{N}{2}-1}(1)} dx \nonumber \\
 &=&\int\limits_0^{\pi} \mbox{exp}\left(\sum\limits^n_{k=1} 
\beta_k \cos^k  
\left(t\right)\right)\sin^{N-2}\left(t\right)\frac{C_l^{\frac{N}{2}-1} 
(\cos\left(t\right))}{C_l^{\frac{N}{2}-1}(1)} dt
\label{EWl}
\end{eqnarray}
with the substitution 
$\mbox{\boldmath$\sigma$}\cdot\mbox{\boldmath$\sigma'$}=:x=\cos(t)$ and $l\in
\N_0$ ($n:=2$ : mixed isovector/isotensor--model); for
$N=2$, we use the {\it Chebychev--polynomials of the first kind} :
$T_0(x)=1$ and
$T_l(x)=\frac{l}{2}\lim\limits_{N\downarrow 2}\frac{2}{N-2}
C_l^{\frac{N}{2}-1}(x)$, for $l\ge 1$.

For the case of nonnegative parameters $\beta_k$ the transfer matrix is a 
positive operator due to reflection positivity \cite{FILS,OS}.
Therefore it is possible to define `masses' in terms of the
normalized eigenvalues
$\tilde{\lambda}_l(\mbox{\boldmath$\beta$})=\lambda_l\lambda_0^{-1}, l\ge 1$,
as
\begin{equation}
m_l(\mbox{\boldmath$\beta$}):=\log\left(\frac{\lambda_0(\mbox{
\boldmath$\beta$})}{\lambda_l(\mbox{\boldmath$\beta$})}\right)=-\log\left(
\tilde{\lambda}_l(\mbox{\boldmath$\beta$})\right).
\label{mass}
\end{equation}

Incidentally,
there is a certain `gauge' symmetry in this action : changing the sign of
all $\beta_k, k$ odd, can be compensated by the substitution $y:=-x$
in the {\it Gegenbauer--polynomials}; this in turn can be achieved by 
multiplying $\mbox{\boldmath$\sigma$}_x$ by $(-1)^x$, which can be 
considered as a gauge transformation not affecting the physics.

We are now going to examine where this model becomes critical and has a
(well--defined) continuum limit; that means that {\it all} masses have
to vanish. In one dimension, this requires that
at least one of the parameters goes to infinity.
Finally we define the ratio of the masses ({\ref{mass}) as
\begin{equation}
{\cal R}_l(\mbox{\boldmath$\beta$}):=\frac{m_l(\mbox{\boldmath$\beta$})}{m_1(
\mbox{\boldmath$\beta$})}
\label{ratio}
\end{equation}
with $m_1(\mbox{\boldmath$\beta$})$ as the reference mass.

\section[Main result : the hypersurfaces for the new universality classes]
        {Main result : the hypersurfaces for the new universality classes
         \label{MainResult}}

In this section we will show how to get (in principle) all
normalized eigenvalues
$\tilde{\lambda}_l:=\lambda_l\lambda_0^{-1}$ from $\lambda_0$.
Because of the impossibility of an exact analytic formula in the
general $n$--parameter case, we use the generalized {\it Laplace--method}  of 
asymptotic expansion techniques \cite{Copson,Murray} to evaluate the
leading term(s) of $\lambda_0$, and thereby also for all
$\tilde\lambda_l$.

Using (\ref{EWl}) it can be seen that the eigenvalues $\tilde\lambda_l$ 
are obtained from $\lambda_0$ as follows:
\begin{equation}
\tilde{\lambda}_l(\mbox{\boldmath$\beta$})=
\frac{2^l}{\Gamma\left(\frac{N}{2}-1\right)C_l^
{\frac{N}{2}-1}\left(1\right)}\sum\limits^
{\left[\frac{l}{2}\right]}_{m=0} \frac{(-1)^m 
\Gamma\left(\frac{N}{2}-1+l-m\right)}{m!\, (l-2m)!\, 
2^{2m}}\frac{\partial}{\partial\beta_{l-2m}}
\left(\log\left(\lambda_0(\mbox{\boldmath$\beta$})\right)\right) 
\quad (N\ge 3)
\label{asympEWlN}
\end{equation}
\begin{equation}
\tilde{\lambda}_l(\mbox{\boldmath$\beta$})=2^l\frac{l}{2}
\sum\limits^{\left[\frac{l}{2}\right]}_{m=0} 
\frac{(-1)^m (l-m-1)!}{m!\,(l-2m)!\,2^{2m}}
\frac{\partial}{\partial\beta_{l-2m}}
\left(\log\left(\lambda_0(\mbox{\boldmath$\beta$})\right)\right) 
\quad (N=2)
\label{asympEWl2}
\end{equation}
with $\frac{\partial}{\partial\beta_0}
\left(\log\left(\lambda_0(\mbox{\boldmath$\beta$})
\right)\right)\equiv 1$.
A priori these two equations are valid for $1\le l\le n$, but we can
apply them for all $l\ge 1$ by the following trick :
First notice that $n$ can be arbitrarily large. We use this fact to
modify the action by introducing additional couplings $\beta_r$ for 
all $r\leq l$ (if not already present); then we take the required 
derivatives and finally set the parameters not appearing in the 
action equal to 0.

Now we turn to the problem of obtaining an asymptotic expansion for
$\lambda_0$.
Let us define first some abbreviations for the expressions in (\ref{EWl})
for the case of $\lambda_0$ (note that
$\frac{C_0^{\frac{N}{2}-1}(x)}{C_0^{\frac{N}{2}-1}(1)}\equiv 1$ ) :
\begin{equation}
f(x):=\sum\limits^n_{k=1} \beta_k x^k,\qquad g(x):=(1-x^2)^{\frac{N-3}{2}}
\label{xfunc}
\end{equation}
\begin{equation}
F(t):=\sum\limits^n_{k=1} \beta_k \cos^k(t),\qquad
G(t):=\sin^{N-2}(t).
\label{tfunc}
\end{equation}
Laplace's method for asymptotic expansion requires the knowledge 
of the absolute maxima of $f(x)$, respectively $F(t)$, in the 
corresponding interval. 
A maximum of $f$ at the point $x_0$ is said to be of order $j$, if 
$f^{(r)}(x_0)=0$ for $r=0,1,\ldots,j-1$ and $f^{(j)}(x_0)\neq 0$. An internal 
maximum is thus at least of order 2, whereas a boundary maximum can have any 
order $j\geq 1$.
Any maximum that contributes to the leading 
term is called a {\it principal maximum}.
We will now consider the simplest case and show in the following section
that these restrictions are unimportant. --- The simplest (but not
trivial) case is given by sending
$|\!|\mbox{\boldmath$\beta$}|\!|\to\infty$ (with any norm 
$|\!|\cdot|\!|$) in such a way that
\begin{equation}
f'(-1)\to -\infty \quad \mbox{and}\quad f'(1)\to\infty\qquad
\mbox{respectively}\qquad F''(\pi)\to -\infty
\quad \mbox{and}\quad F''(0)\to -\infty.
\label{simplecase}
\end{equation}
Whereas at $x=1\, (t=0)$ there is always a (principal) maximum, this is
not a necessity for $x=-1\, (t=\pi)$, but we would like to treat these both 
cases together. The leading term for $\lambda_0$ reads therefore for
$N\in\N\backslash\{1\}$ :
\begin{eqnarray}
\lambda_0(\mbox{\boldmath$\beta$})&= &\frac{\Gamma\left(\frac{N-1}{2}
\right)}{2^{\frac{3-N}{2}}}\frac{G^{\left({N-2}\right)}\left(0\right)}
{\left(N-2\right)!}\frac{\mbox{exp}\left(\sum\limits^n_{k=1} 
\beta_k\right)}{\left((1+\epsilon)\sum\limits^n_{k=1}
k\beta_k\right)^{\frac{N-1}{2}}} \nonumber \\
 & & 
\left(1+\left(\frac{(1+\epsilon)\sum\limits^n_{k=1} k\beta_k}
{(1+\eta)\sum\limits^n_{k=1} (-1)^k k\beta_k}\right)^{\frac{N-1}{2}}
\mbox{exp}\left(-2\sum\limits^{\left[\frac{n+1}{2}\right]}_{k=1} 
\beta_{2k-1}\right)\right)
\label{asympEW0}
\end{eqnarray}
where $\epsilon$ stands for a correction 
$O\left(\left(\sum\limits_{k=1}^n k\beta_k\right)^{-1}\right)$ 
and $\eta$ for 
$O\left(\left(\sum\limits_{k=1}^n (-1)^k k\beta_k\right)^{-1}\right)$.
Due to (\ref{tfunc}) we have (for
$N\ge 3$)
\begin{equation}
G^{\left(N-2\right)}(t)=\partial_t^{N-2}(\sin^{N-2}(t))=\partial_t^{N-2}(
t^{N-2}+O(t^N))=(N-2)! +O(t^2),
\label{Gvalue}
\end{equation}
so that $\frac{G^{\left(N-2\right)}(t_0)}{\left(N-2\right)!}=1$ for
$t_0=0$ and, by way of the transformation  $t\to\pi-t$,
also for $t_0=\pi$.

From these formulae it follows, as we will show below, that
for all $N\ge 2$
\begin{eqnarray}
\tilde{\lambda}_l(\mbox{\boldmath$\beta$})&\sim &
1-\frac{l(l+N-2)}{2}\frac{1}{\sum\limits^n_{k=1} k\beta_k} \nonumber \\
& & -\, \left(1-(-1)^l\right)\, \mbox{exp}\left(-2\sum\limits^{\left[\frac{n+1}
{2}
\right]}_{k=1}\beta_{2k-1}\right)
\left(\frac{\sum\limits_{k=1}^n k\beta_k}
{\sum\limits_{k=1}^n (-1)^k k\beta_k}\right)^{\frac{N-1}{2}}
\nonumber \\
& & +\, O\left(\left(\sum\limits_{k=1}^n k\beta_k\right)^{-2}\right)
\label{asympEWl}
\end{eqnarray}
if at least one of the $\beta_{2k-1}\rightarrow\infty$. In this case
all $\tilde{\lambda}_l$ tend to 1, i.\ e.\ all masses vanish,
so that the model becomes critical. After presenting the proof to 
(\ref{asympEWl}), we will turn back to this point to examine the 
opposite case, in which all $\beta_k, k$ odd,
remain bounded from above.

The proof to (\ref{asympEWl}) requires two steps : first, we will show that 
the essential coefficient of the `power' term is  the eigenvalue of
$\Delta_{LB}$, then we will examine the exponential term.

The eigenvalue of $\Delta_{LB}$ arises from the following identities,
(\ref{EWLBN}) below for
$N\ge 3$ and (\ref{EWLB2}) below for $N=2$, valid in each case for $l\in\N_0$ :
\begin{equation}
\frac{l(l+N-2)}{N-1}=\frac{2^l}
{\Gamma\left(\frac{N}{2}-1\right)C_l^
{\frac{N}{2}-1}\left(1\right)}\sum\limits^
{\left[\frac{l}{2}\right]}_{m=0} 
\frac{(-1)^m (l-2m)\Gamma\left(\frac{N}{2}-1+l-m\right)}{m!\, (l-2m)!\, 2^{2m}}
\label{EWLBN}
\end{equation}
and in the limit $N\downarrow 2$
\begin{equation}
l^2=2^l\frac{l}{2}\sum\limits^{\left[\frac{l}{2}\right]}_{m=0} \frac{(-1)^m 
(l-2m)(l-m-1)!\,}{m!\,(l-2m)!\,2^{2m}}
\label{EWLB2}
\end{equation}
where $\frac{l}{2}$ is the normalization factor coming from 
$T_l(x)=\frac{l}{2}\lim\limits_{N\downarrow 2}\frac{2}{N-2}
C_l^{\frac{N}{2}-1}(x), l\ge 1$, with $T_l$ denoting the
{\it Chebychev--polynomials of the first kind} \cite{Magnus}.

The strategy for the proof of (\ref{EWLBN}) and (\ref{EWLB2})
is to apply the following formulae for $r=1$ and $y=\frac{1}{2}$
\cite{Magnus,Prudnikov1}
\begin{eqnarray}
\sum\limits_{m=0}^{\left[\frac{l}{2}\right]} 
\frac{(-1)^m \Gamma\left(\frac{N}{2}-1+l-m\right)}
{m!\,(l-2m)!}y^{2m}&=&\Gamma\left(\frac{N}{2}-1\right)y^l 
C_l^{\frac{N}{2}-1}\left(\frac{1}{2y}\right), \nonumber \\
\frac{d^r}{dy^r}C_l^{\frac{N}{2}-1}(y)&=& 2^r 
\left(\frac{N}{2}-1\right)_r C^{\frac{N}{2}-1+r}_{l-r}(y)
\label{prN}
\end{eqnarray}
\begin{equation}
\sum\limits_{m=0}^{\left[\frac{l}{2}\right]} \frac{(-1)^m (l-m-1)!}{m!\,
(l-2m)!}y^{2m}=\frac{2}{l}y^l T_l\left(\frac{1}{2y}\right),\quad 
\frac{d^r}{dy^r}T_l(y)=2^{r-1}\Gamma(r)lC^r_{l-r}(y).
\label{pr2}
\end{equation}
In addition we need the identity
\begin{equation}
C_l^{\frac{N}{2}-1}(1)={N+l-3\choose l}
\label{JacobiNorm}
\end{equation}
which arises from the definition of the
{\it Jacobi--polynomials} \cite{Magnus}.
Because of the triviality of the identities (\ref{EWLBN}) and
especially in (\ref{EWLB2}) for $l=0$,
we restrict ourselves to $l\ge 1$ :
\begin{eqnarray}
\lefteqn{\left.\left(-y^{l+1}\Gamma\left(\frac{N}{2}-1\right)\frac{d}{dy}C_l^{
\frac{N}{2}-1}\left(\frac{1}{2y}\right)\right)\right|_{y=\frac{1}{2}}}
\hspace{2cm}
\nonumber\\
 &=& 
-\left.\left(y^{l+1}\Gamma\left(\frac{N}{2}-1\right)2\left(\frac{N}{2}-1\right)
C_{l-1}^{\frac{N}{2}}\left(\frac{1}{2y}\right)\left(-\frac{1}{2y^2}\right)
\right)\right|_{y=\frac{1}{2}} \nonumber\\
 & =& \frac{2}{2^l}\left(\frac{N}{2}-1\right)\Gamma\left(\frac{N}{2}-1\right)
C_{l-1}^\frac{N}{2}(1) \nonumber\\
 &=&\frac{N-2}{2^l}\Gamma\left(\frac{N}{2}-1\right){N+l-2 \choose l-1} 
\nonumber\\
 & =& \frac{\Gamma\left(\frac{N}{2}-1\right)}{2^l}(N-2)\frac{\left(N+l-
2\right)!}{\left(l-1\right)!\,\left(N-1\right)!} \nonumber\\
 & =& \frac{\Gamma\left(\frac{N}{2}-1\right)}{2^l}\frac{\left(N+l-3\right)!}
{l!\,\left(N-3\right)!}\frac{N-2}{N-2}\frac{N+l-2}{N-1}l \nonumber\\
 &=&\frac{\Gamma\left(\frac{N}{2}-1\right)}{2^l}C_l^{\frac{N}{2}-1}(1)\frac{l
\left(l+N-2\right)}{N-1}
\end{eqnarray}
and
\begin{eqnarray}
\left.\left(-y^{l+1}\frac{2}{l}\frac{d}{dy}T_l\left(\frac{1}{2y}\right)\right)
\right|_{y=\frac{1}{2}}&=&
-\left.\left(y^{l+1}\frac{2}{l}lC_{l-1}^1\left(\frac{1}{2y}\right)\left(-
\frac{1}{2y^2}\right)\right)\right|_{y=\frac{1}{2}} \nonumber \\
 & =& \frac{2}{2^l}C_{l-1}^1(1) 
=\frac{2}{2^l}{l \choose l-1}=\frac{l}{2^{l-1}}=
\frac{l^2}{2^l}\frac{2}{l}.
\end{eqnarray}
So far, the first two terms of $\tilde{\lambda}_l$ (\ref{asympEWl}) are
determined. Because in the representation of (\ref{asympEW0})
the argument of the exponential term involves only the $\beta_k, k$ odd,
we distinguish the derivatives of this term 
with respect to $\beta_k$, $k$ even and odd, and denote them
by $\tilde{\partial}_{j,\mbox{{\scriptsize even}}}$ and
   $\tilde{\partial}_{j,\mbox{{\scriptsize odd}}}$ for $1\le j\le n$.
Again, we consider only the leading term.
\begin{eqnarray}
\tilde{\partial}_{j,\mbox{\scriptsize even}}& \sim &\frac{\frac{N-1}{2}
\left(\frac{\sum\limits_{k=1}^n k\beta_k}{\sum
\limits_{k=1}^n (-1)^k k\beta_k}\right)^{\frac{N-3}{2}}
\mbox{exp}\left(-2\sum\limits_{k=1}^
{\left[\frac{n+1}{2}\right]} \beta_{2k-1}\right)
\left(\frac{\displaystyle j}{\sum\limits_{k=1}^n (-1)^k k\beta_k}-
\frac{(-1)^j j\sum\limits_{k=1}^n k\beta_k}
{\left(\sum\limits_{k=1}^n (-1)^k k\beta_k\right)^2}
\right)}{1+\left(\frac{\sum\limits_{k=1}^n k\beta_k}
{\sum\limits_{k=1}^n (-1)^k k\beta_k}\right)^{\frac{N-1}{2}}
\mbox{exp}\left(-2\sum\limits_{k=1}^{\left[\frac{n+1}{2}\right]} 
\beta_{2k-1}\right)} \nonumber \\
 &\sim &-\, \mbox{exp}\left(-2\sum\limits_{k=1}^{\left[\frac{n+1}{2}\right]}
 \beta_{2k-1}\right)\frac{(N-1)j\sum\limits_{k=1}^{\left[\frac{n+1}{2}\right]} 
(2k-1)\beta_{2k-1}}{\left(\sum\limits_{k=1}^n (-1)^k k\beta_k\right)^2}
\left(\frac{\sum\limits_{k=1}^n k\beta_k}{\sum
\limits_{k=1}^n (-1)^k k\beta_k}\right)^{\frac{N-3}{2}}
\label{derieven}
\end{eqnarray}
Because one of the odd $\beta_k\rightarrow\infty$, this contribution is of
order exponential times power and therefore  subleading, compared
to the first two terms in (\ref{asympEWl}).
\begin{eqnarray}
\tilde{\partial}_{j,\mbox{\scriptsize odd}}&\sim &\tilde{\partial}_{j,
\mbox{\scriptsize even}}
+\frac{\mbox{exp}\left(-2\sum\limits_{k=1}^{\left[\frac{n+1}{2}\right]} 
\beta_{2k-1}\right)\left(-2\left(\frac{\sum\limits_{k=1}^n k\beta_k}
{\sum\limits_{k=1}^n (-1)^k k\beta_k}\right)^{\frac{N-1}{2}}\right)}
{1+\left(\frac{\sum\limits_{k=1}^n k\beta_k}{\sum
\limits_{k=1}^n (-1)^k k\beta_k}\right)^{\frac{N-1}{2}}
\mbox{exp}\left(-2\sum\limits_{k=1}^{\left[\frac{n+1}{2}\right]} 
\beta_{2k-1}\right)}\nonumber \\
&\sim & -2\,\mbox{exp}\left(-2\sum\limits_{k=1}
^{\left[\frac{n+1}{2}\right]} \beta_{2k-1}\right)
\left(\frac{\sum\limits_{k=1}^n k\beta_k}
{\sum\limits_{k=1}^n (-1)^k k\beta_k}\right)^{\frac{N-1}{2}}
\label{deriodd}
\end{eqnarray}
We remark that the {\it first} line of (\ref{derieven}) as well as
(\ref{deriodd}) is valid in general (independent of the condition
$\beta_{2k-1}\to\infty$).

If we look at eq.\ (\ref{asympEWl}), we see that three different ways
of sending $|\!|\mbox{\boldmath$\beta$}|\!|\to\infty$
have to be distinguished, depending on the relative importance of the 
second and third terms:
\begin{eqnarray}
\lefteqn{\frac{\sum\limits_{k=1}^n k\beta_k}
{\mbox{exp}\left(2\sum\limits_{k=1}^{\left[\frac{n+1}{2}\right]}
\beta_{2k-1}\right)}
\left(\frac{\sum\limits_{k=1}^n (-1)^k k\beta_k}{\sum\limits_{k=1}^n k\beta_k}
\right)^{\frac{N-1}{2}}}\hspace{1cm}\nonumber \\
&=&\left(\frac{\sum\limits_{k=1}^{\left[\frac{n+1}{2}\right]}(2k-1)\beta_{2k-1}}
{\mbox{exp}\left(2\sum\limits_{k=1}^{\left[\frac{n+1}{2}\right]}
\beta_{2k-1}\right)}+\frac{2\sum\limits_{k=1}^{\left[\frac{n}{2}\right]}k
\beta_{2k}}{\mbox{exp}\left(2\sum\limits_{k=1}^{\left[\frac{n+1}{2}\right]}
\beta_{2k-1}\right)}\right)
\left(\frac{\sum\limits_{k=1}^n (-1)^k k\beta_k}{\sum\limits_{k=1}^n k\beta_k}
\right)^{\frac{N-1}{2}}\nonumber \\
&\stackrel{\left|\!\left|\mbox{\boldmath$\beta$}\right|\!
\right|\to\infty}{\longrightarrow}&
\left\{\begin{array}{c@{\quad:\quad\mbox{Case }}l}
0 &\mbox{(I)}\\
c\in ]0,\infty[ &\mbox{(II)}\\
\infty &\mbox{(III)}
\end{array}\right\}.
\label{cases}
\end{eqnarray}
Here the first summand vanishes in the limit 
$|\!|\mbox{\boldmath$\beta$}|\!|\to \infty$ 
because of the assumption that at least one of the $\beta_{2k-1}\to\infty$. 
The quotient of the linear forms in $\beta_k$ converges always to a value of 
the interval $]0,1]$ (this follows from condition (\ref{simplecase}) in 
connection with (\ref{asympEW0}) and (\ref{Gvalue})).

We are interested in the limit of the mass ratios
${\cal R}_l$ (see (\ref{ratio})) in the three cases just defined. By
expanding the logarithms of the eigenvalues using (\ref{asympEWl}) we obtain
\begin{equation}
{\cal R}_l=\frac{m_l}{m_1}=\lim_{|\!|\mbox{\scriptsize\boldmath$\beta$}|\!|
\rightarrow
\infty}\frac{-\log(\tilde{\lambda}_l(\mbox{\boldmath$\beta$}))}
{-\log(\tilde{\lambda}_1(\mbox{\boldmath$\beta$}))} \nonumber\\
=\left\{\begin{array}{c@{\quad:\quad}l}
\frac{l\left(l+N-2\right)}{N-1} & \mbox{Case (I)} \\
\frac{\frac{l\left(l+N-2\right)}{N-1}+\frac{4}
{N-1}c\frac{1-(-1)^l}{2}}{1+\frac{4}{N-1}c} & \mbox{Case (II)}  \\
\frac{1-(-1)^l}{2} & \mbox{Case (III)}
\end{array}\right\}.
\label{ratioresult}
\end{equation}

The result of (\ref{ratioresult}) can be interpreted as follows : 
In Case (I) the ratio is in the class of the $O(N)$--{\it vector--model},
in Case (III) in the class of the $R P^{N-1}$--{\it model},
whereas in Case (II) we have found {\it new universality classes} lying
between them, parametrized by $c$. 
By the way, if we set all $\beta_k=0$ except of $\beta_1$, 
we recognize the pure $O(N)$--vector--model which belongs to
Case (I).

We can summarize the main result of this paper as follows: the new
universality classes are obtained by sending
$|\!|\mbox{\boldmath$\beta$}|\!|\to\infty$
in such a way that
\begin{equation}
\frac{\sum\limits_{k=1}^n k\beta_k}
{\mbox{exp}\left(2\sum\limits_{k=1}^{\left[\frac{n+1}{2}\right]}
\beta_{2k-1}\right)} 
\left(\frac{\sum\limits_{k=1}^n (-1)^k k\beta_k}{\sum\limits_{k=1}^n k\beta_k}
\right)^{\frac{N-1}{2}}
\to c\in ]0,\infty[
\end{equation}
or equivalently for $|\!|\mbox{\boldmath$\beta$}|\!|\to\infty$
on the hypersurfaces
\begin{equation}
2\sum\limits_{k=1}^{\left[\frac{n}{2}\right]} 
k\beta_{2k}=c\,\mbox{exp}\left(2\sum\limits_
{k=1}^{\left[\frac{n+1}{2}\right]}\beta_{2k-1}\right)+
h\left(\mbox{\boldmath$\beta$}\right),
\qquad c\in ]0,\infty[,
\label{hypersurfaces}
\end{equation}
with
\begin{equation}
\lim_{|\!|\mbox{\scriptsize\boldmath$\beta$}|\!|\to\infty}h\left(
\mbox{\boldmath$\beta$}\right)\exp\left(-2\sum
\limits_{k=1}^{\left[\frac{n+1}{2}\right]}\beta_{2k-1}\right)=0.
\end{equation}
Of course, what matters is only the asymptotic behaviour of those
hypersurfaces which is independent of the function $h$.

Let us return to (\ref{asympEWl}) where we had assumed that at
least one of the $\beta_k, k$ odd, will go to infinity. The converse is that 
all of them are bounded (from above). Then the contribution from 
$\tilde{\partial}_{j,\mbox{\scriptsize even}}$ (\ref{derieven}) 
has purely power character and moreover
is equal to the term whose coefficient is
the eigenvalue of $\Delta_{LB}$.  For
$\tilde{\partial}_{j,\mbox{\scriptsize odd}}$ (\ref{deriodd}) the additional 
term tends to a constant. That means in 
the limit $|\!|\mbox{\boldmath$\beta$}|\!|\rightarrow\infty$ 
(here at least one of the $\beta_k,k $ even, has to go
to infinity), only the $\tilde{\lambda}_l, l$
even, tend to 1, whereas
\begin{equation}
\tilde{\lambda}_{2r+1}\rightarrow 1+\frac{-2\,\mbox{exp}\left(-2\sum\limits_{
k=1}^{\left[\frac{n+1}{2}\right]}\beta_{2k-1}\right)}{1+\mbox{exp}\left(-2\sum
\limits_{k=1}^{\left[\frac{n+1}{2}\right]}\beta_{2k-1}\right)}=\mbox{th}\left(
\sum\limits_{k=1}^{\left[\frac{n+1}{2}\right]}\beta_{2k-1}\right)<1.
\end{equation}
Therefore, the model will not become critical in this case.
Nevertheless, we can consider the ratio (note that there is no
$N$--dependence)
\begin{equation}
{\cal R}_l=\frac{m_l}{m_1}\to \frac{1-(-1)^l}{2}.
\end{equation}
So we end up in the universality class of the pure
$R P^{N-1}$--model, which is the special case in which
all $\beta_k =0$ except of $\beta_2$.

In the following section we will see that this picture remains valid in
the general case of the $n$--parameter model, defined by (\ref{EWl}).

\section[Generalization and continuum limit]
        {Generalization and continuum limit \label{contlim}}

In this section we drop the two restrictions on the action made in section 
\ref{MainResult} and discuss the consequences for a continuum limit.

Firstly, we want to point out that the one restriction made in the beginning 
of the last section, namely that $f'(-1)\to-\infty$ and $f'(1)\to\infty$, is 
irrelevant. Since $f'(1)=\sum\limits_{k=1}^nk\beta_k$ and we are interested 
in the limit $|\!|\beta |\!|\to\infty$, the second condition is obviously 
automatically fulfilled. Let us assume that the first condition is not
satisfied, i.e. $f'(-1)$ is bounded from below. Since we only have to 
consider the case that asymptotically $f$ has a maximum at $x=-1$, we 
may also assume that it is bounded from above, i.e. we have
\begin{equation}
|f'(-1)|=\left|\sum\limits_{k=1}^n (-1)^{k-1}k\beta_k\right|=
\left|\, 2\sum\limits_{k=1}^{\left[\frac{n}{2}\right]}  k\beta_{2k}-
\sum\limits_{k=1}^{\left[\frac{n+1}{2}\right]} (2k-1)\beta_{2k-1}\right|
\le K<\infty.
\label{constraint}
\end{equation}
If now all $\beta_{2k-1}$ as well as all $\beta_{2k}$ remain
bounded, no statement about any asymptotic behaviour is possible. 
Otherwise, both sums above in $\beta_{2k-1}$ and $\beta_{2k}$ have to go 
to infinity. But in this case, using (\ref{constraint}), we get from 
\begin{equation}
e^{f\left(-1\right)-f\left(1\right)}=
\mbox{exp}\left(-2\sum\limits_{k=1}^
{\left[\frac{n+1}{2}\right]}\beta_{2k-1}\right)\le e^{-{K\over n}}\,
\mbox{exp}\left(-{1\over n}\sum\limits_{k=1}^n k\beta_{k}\right)
\end{equation}
that $e^{f\left(-1\right)}$ is exponentially suppressed in {\it all}
$\beta_{2k-1}$  as well as in {\it all} $\beta_{2k}$. Therefore, this 
corresponds to the Case (I) of (\ref{ratioresult}).

Next, we allow (principal) maxima lying inside of the interval :
$x_0\in]-1,1[$, respectively $t_0\in]0,\pi[$. First of all, we want to remark 
that such an internal maximum can appear for
$f$, defined in (\ref{xfunc}), only for $n\ge 3$. As mentioned there, at
$x=1$, we have always a principal maximum for the action in
question. So,
if there is (at least) one internal maximum at $x_0$, notice
that in (\ref{asympEW0}) the exponential term in the 
parenthesis would be replaced by
$\mbox{exp}\left(-\sum\limits_{k=1}^n \left(1-x_0^k\right)\beta_k\right)$.
This means that $e^{f\left(x_0\right)}$ is 
exponentially suppressed in
each $\beta_k,1\le k\le n$, and therefore in the asymptotic expansion
the contribution from $x_0$ would be subleading. This completes
the proof of the statement that the restrictions made in section
\ref{MainResult} are unimportant.

Let us now discuss the continuum limit. This and the problem
for principal maxima of different orders is discussed in some detail in 
\cite{Sokal} for the case of a family with at most two parameters and 
a general action. They find that if the only maximum is at an internal
point of the interval (a situation excluded by our assumption that all
the parameters $\beta_k$ are nonnegative), there exists no continuum 
limit. For $N\ge 3$, it is shown that the normalized eigenvalues
$\tilde{\lambda}_l$ of (\ref{EWl}) cannot tend to 1 (so that the masses 
(\ref{mass}) would vanish), because of the fact that
\begin{equation}
\frac{\left|C_l^{\frac{N}{2}-1}(x_0)\right|}{C_l^{\frac{N}{2}-1}(1)}<1
\qquad \mbox{for }|x_0|<1\quad\mbox{and }l\in\N.
\label{Gegenbineq}
\end{equation}
Unfortunately, this does not cover the special case $N=2$. But we want to give
an alternative (simpler) proof for this fact which is also valid for $N=2$.

We use the following formula for the {\it Gegenbauer--polynomials} in terms of
the {\it Chebychev--polynomials of the first kind} \cite{Bateman1} (valid 
for $N\ge 3$)
\begin{equation}
C_l^{\frac{N}{2}-1}(T_1(x))=\frac{2}{\left(\Gamma\left(\frac{N}{2}-1\right)
\right)^2}\sum\limits_{m=0}^{\left[\frac{l}{2}\right]} \frac{\Gamma\left(\frac{
N}{2}-1+m\right)\,\Gamma\left(\frac{N}{2}-1+l-m\right)}{m!\,(l-m)!}T_{l-2m}(x).
\label{GegChe}
\end{equation}
Note that all coefficients are positive! To prove (\ref{Gegenbineq}),
it suffices to show that for any $t_0\in]0,\pi[$ one of the 
$T_{l-2m}(x_0)$ in (\ref{GegChe}) (with $x_0=\cos(t_0)$) is less than 1. (Of 
course, for $l=1$, we have only one term, $T_1(x)=x\equiv\cos(t)$, whose 
absolute value is always less than 1 for such a $t_0$.)
Assume that $T_l(x_0)=\pm 1$ (only one sign to choose). Without loss of
generality , consider for $l\ge 2$
\begin{equation}
T_{l-2}(x_0)=\cos((l-2)t_0)=\cos(lt_0)\,\cos(2t_0)-\sin(lt_0)\,\sin(2t_0)=
\pm\cos(2t_0)=\pm T_2(x_0)\neq \pm 1.
\end{equation}

For $N=2$, the inequality (\ref{Gegenbineq}) is not true for certain
points $t_0=\pi q, q\in\Q\,\cap \,]0,1[$. But we can repeat the
arguments used above : assume $T_l(x_0)=\pm 1$ (again, $l\ge 2$ and one 
sign to choose); this time, consider
\begin{equation}
T_{l-1}(x_0)=\pm T_1(x_0)\neq \pm 1.
\end{equation}
This means that not all masses would go to 0 if at such a $t_0$ the
function $F$ (see eq.\ (\ref{tfunc})) has a maximum. Consequently, in these
cases, there exists no continuum limit. So, as far as the continuum limit
is concerned, there is nothing special about the case $N=2$.

Our conclusions are in agreement with those of \cite{Sokal}, where
they overlap. Our analysis is, however, more general in one respect :
we allow for arbitrarily many parameters, whereas the authors of
 \cite{Sokal} allow only one or two. On the other hand, the form 
of interactions considered by us is more restricted since we only 
consider polynomials in the scalar products of two neighbouring spins 
with nonnegative coefficients.

It is easy to see that the continuum limits obtained in the new
universality classes correspond to quantum--mechanical Hamiltonians
of the form
\begin{equation}
H=-a\Delta_{LB}+bP+\mbox{const}
\label{Ham}
\end{equation}
where $\Delta_{LB}$ is the {\it Laplace--Beltrami--operator} on $S^{N-1}$
and $P$ is the `parity operator' mapping every point of the sphere into its
antipode (note that $P$ is a unitary, self-adjoint involution). Since $P$
corresponds to multiplication by $(-1)^l$ on the eigenspace ${\cal H}_l$
of $\Delta_{LB}$, it is not hard to check that we obtain the mass ratios
of the new universality classes given in (\ref{ratioresult})
by choosing $a=\frac{1}{N-1}$ and $b=-\frac{2c}{N-1}$ (and $\mbox{const}=-b$).

We do not get the more general continuum Hamiltonians discussed
in  \cite{Sokal} because we restricted ourselves to polynomials
with nonnegative coefficients in order to ensure reflection positivity.
It should be noted, however, that the most general quantum--mechanical
Hamiltonian compatible with $O(N)$--invariance is still more general
than the form given in \cite{Sokal} : it is given by
\begin{equation}
H=\sum_{l\geq 1} c_l{\cal P}_l
\label{Hamgen}
\end{equation}
where the ${\cal P}_l$ are the projectors onto the eigenspaces, defined in 
(\ref{projectors}), and $c_l$ arbitrary coefficients.

\section[Conclusions]{Conclusions\label{conclusion}}

In this paper we determined the critical behaviour of the generalized
$O(N)$--$R P^{N-1}$--model with an $n$--parametric action
for the general case in one dimension and for general $N$
using asymptotic expansion techniques. There exist hypersurfaces on
which an infinite number of new universality classes appears, which
can be parametrized
by a variable interpolating between the $O(N)$--vector--model--class
and the $R P^{N-1}$--model--class. For the ratio of the masses, there is a
difference between even and odd masses in form of an additional constant.
We also examined
the continuum limit and gave some relevant additional information
for the case $N=2$.

We found a one--parameter family of universality classes in the
continuum limit, described in eq.\ (\ref{hypersurfaces}), that arises in 
particular for the standard mixed $O(N)$--$RP^{N-1}$ models. These models were
also studied in the papers \cite{NSW} and \cite{Hasenbusch};
these authors concluded, however, that there is only the universality
class corresponding to the standard $O(N)$--model. At least the arguments
of \cite{NSW} that rely on the negligibility of vortices are
clearly applicable in our $1 D$ situation. So how is this apparent
conflict to be resolved ?

The authors of\ \cite{NSW} reach their conclusion by considering
the decay properties
of the correlations of new spin variables, and their claims for those is
certainly correct. But these new spins are {\it nonlocal} functions of 
the original spins, and therefore the behaviour of their correlations
cannot be related in an obvious way to spectral properties of the
transfer matrix. We analyzed directly the spectrum of the transfer
matrix and found that there are indeed new universality classes.

Finally, we would like to make a remark about a possible generalization
of our results to negative values of some parameters. If negative (or 
even complex) values are allowed, we generally lose reflection
positivity and therefore the quantum--mechanical interpretation of
the models. The most general Hamiltonians given in eq. (\ref{Hamgen}) 
can certainly be obtained as continuum limits of such lattice models;
we only have to choose as the transfer matrix ${\cal T}=\exp(-aH)$
and send $a\to 0$. Such transfer matrices correspond, however, to
very involved actions that depend on $a$ (the parameter controlling
the approach to criticality) in a very
nonobvious way. It remains an open question whether the Hamiltonians
(\ref{Hamgen}) can also be obtained by using more `natural' actions.

\vspace{1cm}

{\Large \bf Acknowledgement}

We are grateful to A.\ Sokal for correspondence about \cite{Sokal}.

\begin{thebibliography}{999}
\bibitem{Caracciolo}  S.\ Caracciolo, R.\ G.\ Edwards, A.\ Pelissetto and 
                      A.\ D.\ Sokal, Phys.\ Rev.\ Lett.\ 
                      {\bf 71}, 3906, 1993
\bibitem{NSW}  F.\ Niedermayer, D.--S.\ Shin and P.\ Weisz, {\it On the 
               Question of Universality in} $RP^{N-1}$ {\it and} $O(N)$ {\it
               Lattice Sigma Models}, hep--lat/9507005
\bibitem{Hasenbusch}  M.\ Hasenbusch, $O(N)$ {\it and} $RP^{N-1}$ {\it Models
                      in Two Dimensions}, hep--lat/9507008
\bibitem{Sokal}  A.\ Cucchieri, T.\ Mendes, A.\ Pelissetto and A.\ D.\ Sokal,
                 {\it Continuum Limits and Exact Finite--Size--Scaling 
                 Functions for One--Dimensional} $O(N)$--{\it Invariant
                 Spin Mo\-dels}, hep--lat/9509021
\bibitem{Magnus}  W.\ Magnus, F.\ Oberhettinger and R.\ P.\ Soni, {\it 
                  Formulas and Theorems for  the Special Functions of 
                  Mathematical Physics}, Bd.\ {\bf 52}, Springer--Verlag, 1966
\bibitem{FILS}  J.\ Fr\"ohlich, R.\ Israel, E.\ Lieb und B.\ Simon,
                 Commun.\ Math.\ Phys.\ {\bf 62}, (1978) 1
\bibitem{OS}  K.\ Osterwalder and E.\ Seiler, Ann.\ Phys.\ {\bf 110}, (1978) 440
\bibitem{Copson}  E.\ T.\ Copson, Cambridge Tracts in Mathematics and 
                  Mathematical Physics, No.\ {\bf 55}, 
                 {\it Asymptotic expansion}, Cambridge
                  University Press, Cambridge, 1965
\bibitem{Murray}  J.\ D.\ Murray, {\it Asymptotic analysis}, Clarendon Press,
                  Oxford, 1974
\bibitem{Prudnikov1}  A.\ P.\ Prudnikov, Yu.\ A.\ Brychkov and O.\ I.\ 
                      Marichev, {\it Integrals and Series}, \mbox{Vol.\ {\bf 1}                       :} Elementary Functions, Gordon and Breach Science
                      Publishers, 1983
\bibitem{Bateman1}  A.\ Erd{\'e}lyi, W.\ Magnus, F.\ Oberhettinger and  F.\ G.\
                    Tricomi, Bateman Manuscript Project, {\it Higher
                   Transcendental Functions}, Vol.\ {\bf 1}, McGraw--Hill, 1953
\end{thebibliography}

\end{document}